**Floating Forests: Quantitative Validation of Citizen Science Data Generated From Consensus Classifications**

Isaac S. Rosenthal (Department of Biology, University of Massachusetts Boston, Boston, MA 02125 USA),

Jarrett E.K. Byrnes (Department of Biology, University of Massachusetts Boston, Boston, MA 02125 USA),

Kyle C. Cavanaugh (Department of Geography, University of California, Los Angeles, CA 90095),

Tom W. Bell [2] (Department of Geography, University of California, Los Angeles, CA 90095),

Briana Harder,

Alison J. Haupt (School of Natural Sciences, California State University Monterey Bay, Seaside, CA 93955 USA),

Andrew T.W. Rassweiler (Department of Biological Science, Florida State University, Tallahasse, FL 32306),

Alejandro Pérez-Matus (Estación Costera de Investigaciones Marina, Pontificia Universidad Católica de Chile, Osvaldo Marín 1672, Las Cruces, Comuna El Tabo, V Región, Chile),

Jorge Assis (Center of Marine Sciences, CCMAR- CIMAR, University of Algarve, Faro, Portugal),

Ali Swanson (The Zooniverse, Adler Planetarium, Chicago, IL 60605),

Amy Boyer (The Zooniverse, Adler Planetarium, Chicago, IL 60605),

Adam McMaster (The Zooniverse, Adler Planetarium, Chicago, IL 60605),

Laura Trouille (The Zooniverse, Adler Planetarium, Chicago, IL 60605)

**ABSTRACT**

Large-scale research endeavors can be hindered by logistical constraints limiting the amount of available data. For example, global ecological questions require a global dataset, and traditional sampling protocols are often too inefficient for a small research team to collect an adequate amount of data. Citizen science offers an alternative by crowdsourcing data collection. Despite growing popularity, the community has been slow to embrace it largely due to concerns about quality of data collected by citizen scientists. Using the citizen science project Floating Forests (http://floatingforests.org), we show that consensus classifications made by citizen scientists produce data that is of comparable quality to expert generated classifications. Floating Forests is a web-based project in which citizen scientists view satellite photographs of coastlines and trace the borders of kelp patches. Since launch in 2014, over 7,000 citizen scientists have classified over 750,000 images of kelp forests largely in California and Tasmania. Images are classified by 15 users. We generated consensus classifications by overlaying all citizen classifications and assessed accuracy by comparing to expert classifications. Matthews correlation coefficient (MCC) was calculated for each threshold (1-15), and the threshold with the highest MCC was considered optimal. We showed that optimal user threshold was 4.2 with an MCC of

0.400 (0.023 SE) for Landsats 5 and 7, and a MCC of 0.639 (0.246 SE) for Landsat 8. These results suggest that citizen science data derived from consensus classifications are of comparable accuracy to expert classifications. Citizen science projects should implement methods such as consensus classification in conjunction with a quantitative comparison to expert generated classifications to avoid concerns about data quality.

**Introduction**

Much of the scientific community has been slow to embrace citizen science despite the potential it has to massively increase the scale at which research can be done.(Dickinson, Zuckerberg, & Bonter, 2010) As the scientific community turns its eye towards global issues such as climate change, research is bottlenecked because the requisite amount of data to tackle these questions simply cannot collected by traditional small research teams (Ricciardi, Steiner, Mack, & Simberloff, 2000). Citizen science offers an efficient method to collect a dataset of adequate size to tackle these large-scale questions by crowd-sourcing tasks that would otherwise be prohibitively time consuming (Willett et al., 2013). Citizen science also provides a rare and valuable opportunity for collaboration between researchers and members of the public (Cooper, Dickinson, Phillips, & Bonney, 2007). This is of particular import in ecological research, as the outcome of a study could lead to management decisions with implications for the general public (Lewandowski & Specht, 2015). A disconnect exists between the general public and their understanding of the scientific process that can be lessened by participating in citizen science projects and interacting with researchers on a personal basis (Irwin, 2001). However, to be useful, we need to be assured that citizen science is generating data of sufficient quality to produce meaningful results. This problem is pernicious in scientific circles, leading some professional scientists to look askance at citizen science projects (Bird et al., 2014; Darwall & Dulvy, 1996; Lewandowski & Specht, 2015). Here we present a simple validation method utilizing consensus between multiple citizen scientists to generate high quality data. We demonstrate how, for remote sensing data of giant kelp, it can produce data comparable in accuracy to expert scientists.

*Citizen Science*

Citizen science is not a new concept; there are examples in both astronomy and ornithology as old as the late 18th century (Dickinson et al., 2010; Greenwood, 2007). More recently, several citizen science ornithological studies have become household names, including the Audubon Christmas Bird Count, Cornell's ProjectFeederWatch, NestWatch and eBird, and USGS's Breeding Bird Survey (National Audubon Society, 2018; National Audubon Society & The Cornell Lab of Ornithology, 2018; The Cornell Lab of Ornithology, 2018a, 2018b; USGS Patuxent Wildlife Research Center, 2018). This citizen science approach to collaborative field data collection has bled into other fields of ecology, such as the North American Butterfly Count which began in 1993 and covers the United States, Canada, and Mexico (North American Butterfly Association, 2018). Over the last decade, the power and reach of the internet has led to an explosion of citizen science activity in other fields as well (Cooper et al., 2007; Dickinson et al., 2010; Lepczyk et al., 2009). Galaxy Zoo is an extremely successful example of online citizen science and has resulted in over 50 publications (Zooniverse, 2018). The biomedical field also has found success through projects such as Foldit, in which protein folding has been gamified (Foldit, 2018). This list makes up a tiny fraction of the larger citizen science body, with at least 937 active citizen science projects running as of last assessment (SciStarter, 2018)

Despite its potential as a powerful research tool, the scientific community has been hesitant to embrace citizen science. This reluctance is largely due to a lack of rigorous validation standards which results in many datasets of unknown quality and has led to a general attitude of distrust of citizen scientists' data (Delaney, Sperling, Adams, & Leung, 2008). Despite these concerns, there are several ways to overcome the question of data quality in citizen science. Perhaps the most conventional is to simply ensure a large sample size; more data typically means more precision of estimated population parameters despite increases in variance of the data (Dickinson et al., 2010). This should not be a major obstacle for most citizen science projects, as often the decision to engage in citizen science was made because of the size of the required dataset (Silvertown, 2009). Still, even with a large dataset, scientists are often skeptical without some means of quality control.

Large datasets do not guarantee accurate data (Dickinson et al., 2010). To ensure data quality, data collected by citizens must be compared with data collected by experts. Beyond simply validating a dataset, these comparisons can be invaluable for developing more comprehensive sampling regimes as lessons learned in the validation of a pilot dataset can be applied to the definition of volunteer eligibility and sampling protocols, in turn ensuring higher accuracy for the main body of work (Boudreau & Yan, 2004; De Solla et al., 2005; Delaney et al., 2008).

Many projects, particularly those that are web-based, do not have set requirements for volunteer eligibility. These studies rely on *post hoc* comparisons between citizen scientist and expert classifications to ensure data quality. For example, in Cornell's ProjectFeederWatch over 14,000 citizen scientists contribute to over 5,000,000 individual bird observations on an annual basis. Cornell has developed a semi-automated system in which anomalous observations are flagged and then reviewed by experts before being re-integrated into the dataset (Bonter & Cooper, 2012). This method allows integration of quality-control protocols directly into the data generation pipeline contributes to efficiently processing large amounts of citizen science data without compromising quality. Developing methods to ensure data quality without restricting volunteer eligibility requirements is a priority as web-based citizen science such as becomes increasingly popular (Bonney et al., 2014).

Consensus classification leverages agreement between multiple citizen scientists to improve quality of data provided by citizen scientists of unknown and varied backgrounds and stands in contrast to methods that seek to rely on individuals for quality data (Hutt, Everson, Grant, Love, & Littlejohn, 2013; Swanson et al., 2015). The foundation of consensus classification lies in redundant processing of samples by multiple volunteers. This results in multiple complete sets of classifications that can then be aggregated to suit the researcher's needs. An advantage of consensus classification is that data quality is preserved even if a number of individual citizen scientists are inaccurate (Hutt et al., 2013). This allows projects to tap into many demographics in search of volunteers without concerns about previous experience or a need to judge the abilities of individual citizen scientists.

The consensus approach has proved popular, particularly with online citizen science. Here we demonstrate its efficacy using a citizen science project to detect thirty years of change in the world's kelp forests, Floating Forests (http://floatingforests.org). Floating forests uses consensus classifications to crowdsource detection of giant kelp (*Macrocystis pyrifera*) in Landsat satellite images in an effort to establish a global picture of kelp distribution and health over the last 30 years (Bell, Cavanaugh, & Siegel, 2015; Cavanaugh, Siegel, Kinlan, & Reed, 2010). Traditional field sampling approaches are far too labor intensive to attain the requisite global coverage, and for obvious reasons cannot be performed retroactively. Existing techniques for measuring kelp canopy cover from Landsat satellite images require hundreds of human work-hours to classify even a relatively small geographic region. We show that citizen science utilizing consensus classifications can classify kelp in Landsat images on a large scale with comparable accuracy to expert classifications.

**Methods**

*Landsat Processing*

Full size Landsat scenes were converted into small jpeg subsets that were presented to citizen scientists via the Floating Forests classification interface. For each region (e.g. California and Tasmania) we used NOAA's World Vector Shoreline dataset to identify the path/rows that contained coastline. All available images for each path/row were downloaded from the USGS Landsat archives. Each Landsat image was converted to top of atmosphere reflectance using scene specific bias and gain values, earth-sun distance, and solar zenith angle. We split each Landsat scene into 400 images of equal size along a 20 x 20 grid (~131 $km^2$ per image). Each image subset was displayed with the short-wave infrared band as red, the near infrared band as green, and the red band as blue. The high near infrared reflectance of kelp canopy caused it to stand out as bright green with this band combination.

*Floating Forests*

We evaluated the accuracy of consensus classification using our Floating Forests citizen science website (http://www.floatingforests.org). There were no requirements to participate. All users viewed a brief tutorial which oriented them with the website and provided training on how to identify and classify kelp patches on their first visit. Additionally, a field guide was accessible at any time and contained entries on image features that are commonly confusing. We then showed them randomly selected preprocessed Landsat images and tasked them with tracing the borders of any visible kelp patches using a free-form selection tool (Figure 1). We also asked them to indicate the presence of clouds in the image. If there were any problems with the classification, users were encouraged to tag the image, which would then cross-post it to a talk forum where they could interact directly with researchers. We launched the project in August of 2015 and ran it with imagery from California and Tasmania taken between 1983-2012 until December of 2017. In 2017 the project was relaunched with a different Landsat imagery processing pipeline; the data from Floating Forests 2.0 will not be considered here. The first version of the platform hosted 7,155 users that contributed 758,504 classifications. To aid users, they could also flag images if there is confusion regarding their classification and researchers can address any issues over an online talk forum. Data from Landsat 5, 7, and 8 was used in this analysis which validates citizen scientist classifications in central and southern California.

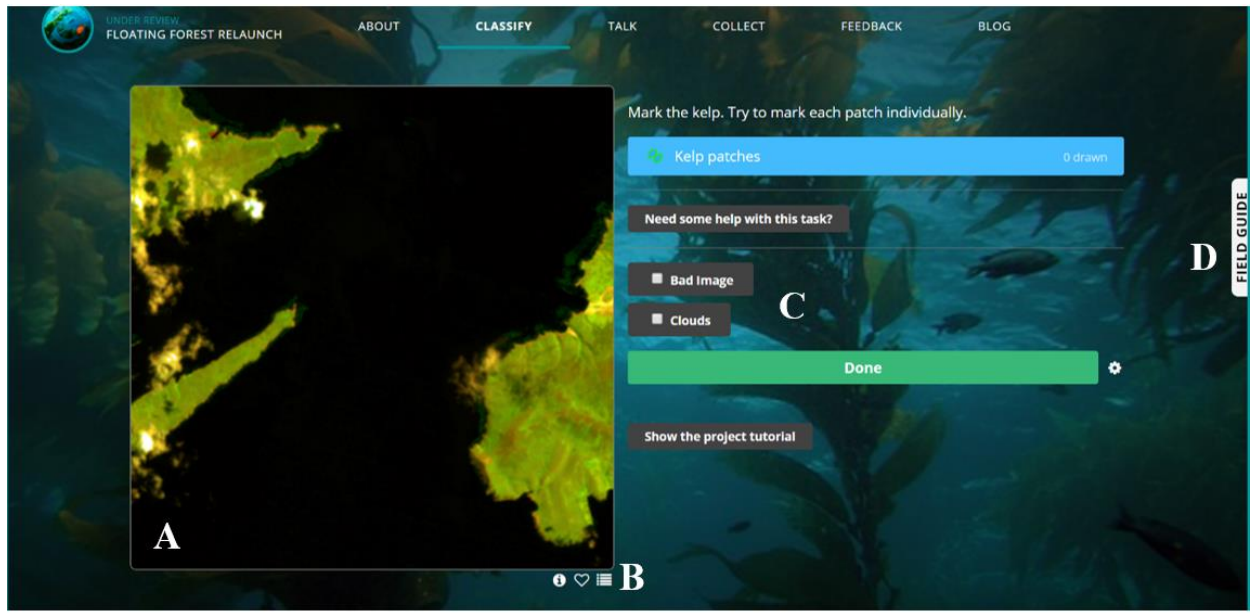

*Figure 1. Classification interface of Floating Forests. Points of interest include the following: A) Classification window. B) Additional image information, including geographic coordinates, Landsat metadata, and a link to view image on Google maps. C) Additional user flags to indicate if an image should be retired as a bad image or contains clouds. D) Field guide containing examples of potentially confusing features that users could encounter.*

*Consensus Classifications*

Floating Forests ensured data quality partially through the use of consensus classifications. All images were classified by at least four users. If any user detected kelp in the image, it was retained in the system until it had been classified by a total of 15 users. However, if no user detected kelp in an image, it was retired and removed from the image pool. Additionally, if an image was flagged by a user as a "bad image", it was dropped from the pool (Figure 4, Figure 5).

*Expert Classifications*

Calibration data for California was obtained from previous work to estimate kelp biomass and patch borders from Landsat photographs (Bell et al., 2015; Cavanaugh, Siegel, Reed, & Dennison, 2011). In short, kelp estimates were derived from the relationship between kelp detected in satellite images and aerial surveys; see (Cavanaugh et al., 2011) for detailed methods.

*Validating Consensus Classifications*

To assess which user threshold produced optimum classifications and to assess the overall quality of classifications, we created a "consensus" dataset by overlaying all user classifications for one image. Each pixel receives a score from 1-15 corresponding to the number of unique "kelp" classifications it received. To score each user threshold (from 1-15 people saying a pixel contained kelp) for each image, we compared the consensus and expert classifications using Matthews Correlation Coefficient (MCC). The user threshold with the highest MCC was considered the optimal user threshold for that image. MCC provides a method to assess the performance of a binary classifier utilizing the confusion matrix for each comparison to produce the following score:

$$\mathbf{MCC} = \frac{\mathbf{TP \times TN - FN \times FP}}{\sqrt{(\mathbf{TP + FN})(\mathbf{TP + FP})(\mathbf{TN + FN})(\mathbf{TN + FP})}}$$

In the equation above, TP and TN refer to true positive and true negative, respectively with FP and FN being false positive and false negative, respectively. For Floating Forests, true positives are pixels correctly classified as kelp, and true negatives are pixels that were correctly not classified as kelp. False positives are pixels that were classified as kelp when in reality they are not, and false negatives were pixels that contained kelp but were not classified. MCC ranges from -1 to 1, with -1 being a completely wrong classifier, 0 representing a classifier being as good as a coin toss, and 1 being a perfect classifier. The can also be interpreted in terms of strength akin to a Pearson Correlation. We used MCC as opposed to other methods to evaluate classifiers (e.g., AUC, or Youden's J) as it works well in cases with low a prevalence of one class (Chicco, 2017). Here we had a low number of kelp pixels in comparison to the total number of pixels in an image (Table 1). In such a scenario, other methods often produce biased results. For example, our levels of specificity (defined as: $SPC = \frac{TN}{(TN+FP)}$) were almost always 1 due to the paucity of kelp relative to other types of pixels in our images. To answer our questions about optimal threshold and overall accuracy, we used polynomial regression to assess the relationship between the optimal user threshold for each subject (i.e., where the highest MCC for that image was obtained) and said optimal MCC. As satellite sensor and season of acquisition can potentially effect user classification accuracy, we included these as predictors of optimal MCC as well as allowing them to interact with user threshold.

*Evaluation of Rejection Rules*

To assess image retirement rules, we used calibration data to summarize the number of expert classified kelp pixels in each image. For retirement rules to be effective they must only drop images with very little kelp or "bad images" which contain incomplete or glitched satellite data. Rejected images that contain a non-zero amount of kelp were visually inspected to confirm that they were accurately flagged as "bad images".

*Evaluation of Bias*

Along with accuracy, it is possible that citizen scientists are more likely to over- or under-classify kelp. To assess whether users were biased we calculated the density of the difference between false negative (indicating over classification) and false positive indicating under-classification. Bias would be indicated if the mean of the density significantly departs from zero.

| User Threshold | User Kelp Pixels | Calibration Kelp Pixels | Total Kelp Pixels | True Positive | True Negative | False Positive | False Negative | MCC |
|---|---|---|---|---|---|---|---|---|
| 1 | 1215 | 505 | 145600 | 505 | 144447 | 710 | 62 | 0.606 |
| 2 | 797 | 489 | 145600 | 489 | 144881 | 308 | 78 | 0.726 |
| 3 | 692 | 472 | 145600 | 472 | 145003 | 220 | 95 | 0.752 |
| 4 | 646 | 467 | 145600 | 467 | 145054 | 179 | 100 | 0.771 |
| 5 | 586 | 459 | 145600 | 459 | 145122 | 127 | 108 | 0.795 |
| 6 | 538 | 447 | 145600 | 447 | 145182 | 91 | 120 | 0.809 |
| 7 | 488 | 425 | 145600 | 425 | 145254 | 63 | 142 | 0.807 |
| 8 | 454 | 403 | 145600 | 403 | 145310 | 51 | 164 | 0.794 |
| 9 | 422 | 376 | 145600 | 376 | 145369 | 46 | 191 | 0.768 |
| 10 | 383 | 347 | 145600 | 347 | 145437 | 36 | 220 | 0.744 |
| 11 | 353 | 328 | 145600 | 328 | 145486 | 25 | 239 | 0.732 |
| 12 | 317 | 296 | 145600 | 296 | 145554 | 21 | 271 | 0.697 |
| 13 | 260 | 251 | 145600 | 251 | 145656 | 9 | 316 | 0.653 |
| 14 | 211 | 207 | 145600 | 207 | 145749 | 4 | 360 | 0.598 |
| 15 | 147 | 147 | 145600 | 147 | 145873 | 0 | 420 | 0.508 |

*Table 1. Sample output from one image demonstrating how confusion matrix and mcc change based on user threshold. In this table, user kelp pixels, calibration kelp pixels, and total kelp pixels refer to number of pixels selected by users, number of kelp pixels in calibration data, and total number of pixels in the image, respectively. Original image can be viewed at https://static.zooniverse.org/www.floatingforests.org/subjects/53e2f4904954734d8b5d1000.jpg. Landsat-8 image courtesy of the U.S. Geological Survey.*

**Results**

In our analysis of optimal MCC, we found a peaked relationship between user threshold and optimal MCC (squared term estimate = -0.006, SE = 0.019, t = 3.060, P <0.001, supplemental table 1). Satellite identity affected MCC, but did not modify the relationship between user threshold and MCC (Table 2). We determined that optimal user threshold was 4.2 with a MCC of 0.400 ± 0.023SE for Landsats 5 and 7, and a MCC of 0.639 ± 0.246 for Landsat 8 (Figure 2).

Rejected images were summarized as a histogram of kelp pixel counts in expert classifications. Figure 3A shows that the majority of rejected images contain little to no kelp. We visually inspected images with a significant amount of expert classified kelp that were still rejected. This inspection showed that these were correctly rejected as bad/glitched images (Figure 3B).

A visual inspection of the false negative to false positive rate indicates no bias in the results; users were not consistently over or under estimating kelp coverage (Figure 4).

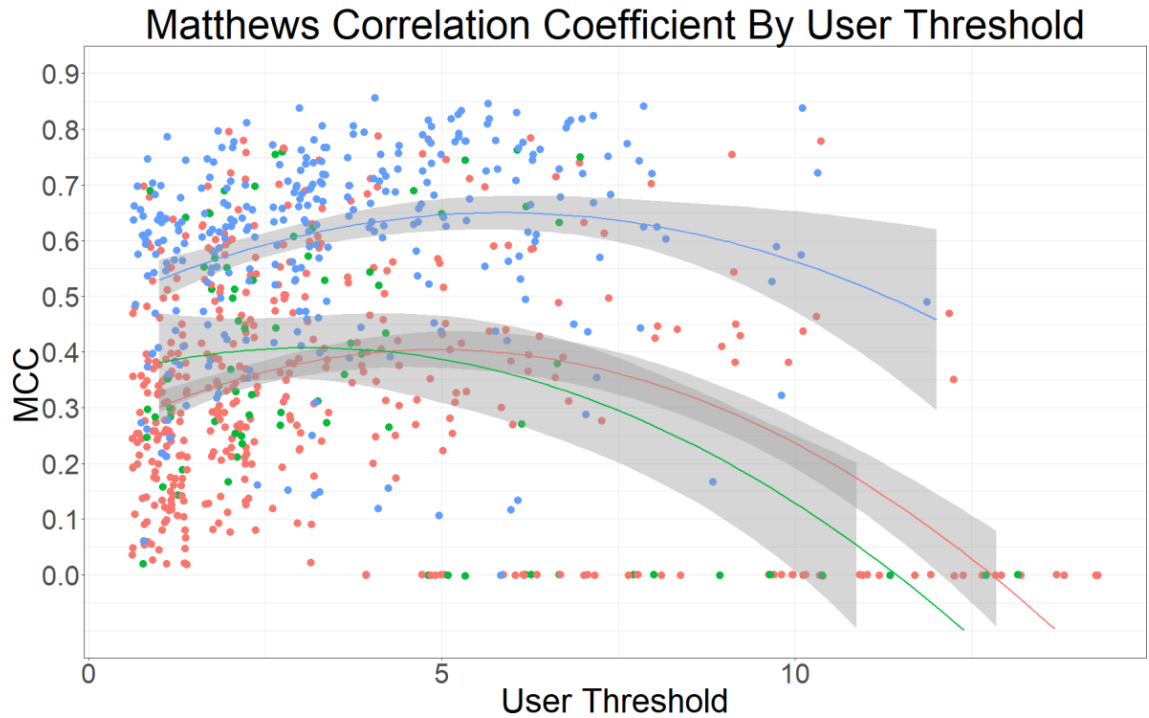

*Figure 2. Model output displaying correllation between user threshold and MCC for each satellite (Landsat 5, 7, and 8). Landsat 5 and 7 did not differ in optimal user threshold (4.2) or MCC (0.400, SE: 0.023). Landsat 8 had the same optimal user threshold (4.2), but had a significantly higher MCC (0.639, SE: 0.246 (Table 2). There was no effect of season on either optimal user threshold or MCC (Table 2).*

| X | Sum. Sq | Df | F | *p*- value |
|---|---|---|---|---|
| User Threshold | 3.74 | 2 | 53.51 | <0.001 |
| Spacecraft ID | 8.25 | 2 | 118.01 | <0.001 |
| Season | 0.11 | 3 | 1.09 | 0.351 |
| User Threshold : Spacecraft ID | 0.28 | 4 | 1.97 | 0.097 |
| User Threshold : Season | 0.15 | 6 | 0.69 | 0.655 |
| Residuals | 26.70 | 764 | | |

*Table 2: F table from ANOVA*

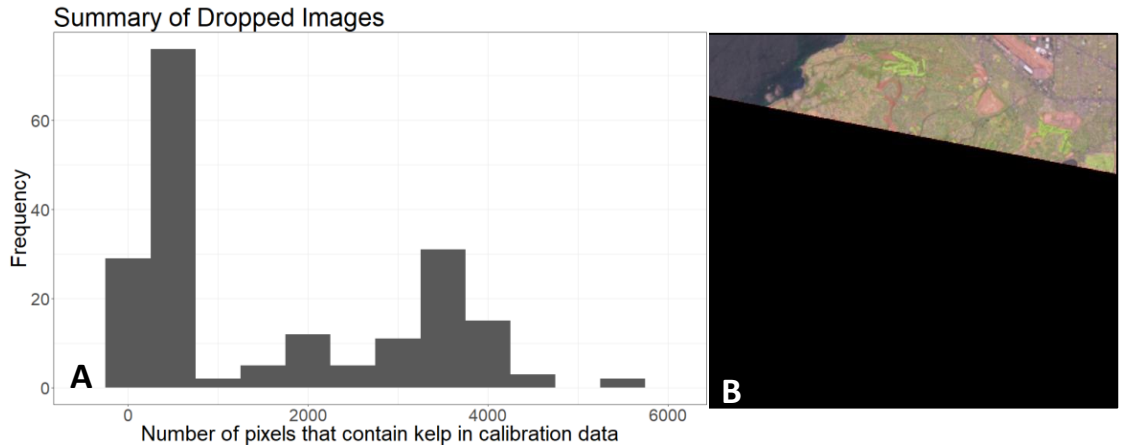

Figure 3. A) Distribution of number of pixels containing kelp per image in expert classifications among subjects retired via user flags. Images with low pixel count were retired due to "no kelp" flags, whereas images with higher pixel counts were retired due to "bad image" flags. B) Subject AKP000imr1 is typical of images retired from classification workflow via a "bad image" user flag. Landsat-5 image courtesy of the U.S. Geological Survey.

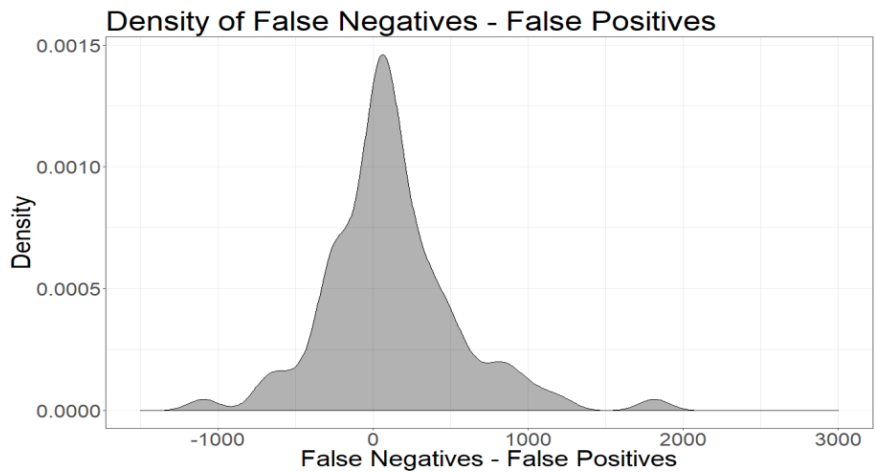

Figure 4. Density of false negatives – false positives. True mean is not significantly different from zero, indicating a lack of bias towards over or underestimation of kelp (t = 1.8323, df = 90, p-value = 0.07021)

## Discussion

Our analysis shows that data derived from citizen science using consensus classifications can be used confidently with comparable accuracy to expert classifications. We found that an optimal consensus threshold of 4.2 users produced an average MCC of 0.400 for Landsats 5 and 7 and an MCC of 0.639 for Landsat 8 (Figure 2). For reference, a MCC of 0.5 would indicate that a classifier is correct 75% of the time (Vihinen, 2012). Further, citizen scientist results are unbiased. This implies that an accurate result can be obtained from a relatively low level of consensus among volunteers, with Landsat 8 producing more accurate classifications than older satellites. The fact that MCC appears to differ between Landsat missions is not unexpected as Landsat 8 produces higher quality images. The consensus approach yields enormous dividends in terms of worries about individual citizen scientists. The accuracy of any one single citizen scientist does not have to be high given that final classifications are determined by consensus. This approach naturally eliminates outliers and, in this example, allowed for highly accurate definition of kelp patch borders (Figure 5).

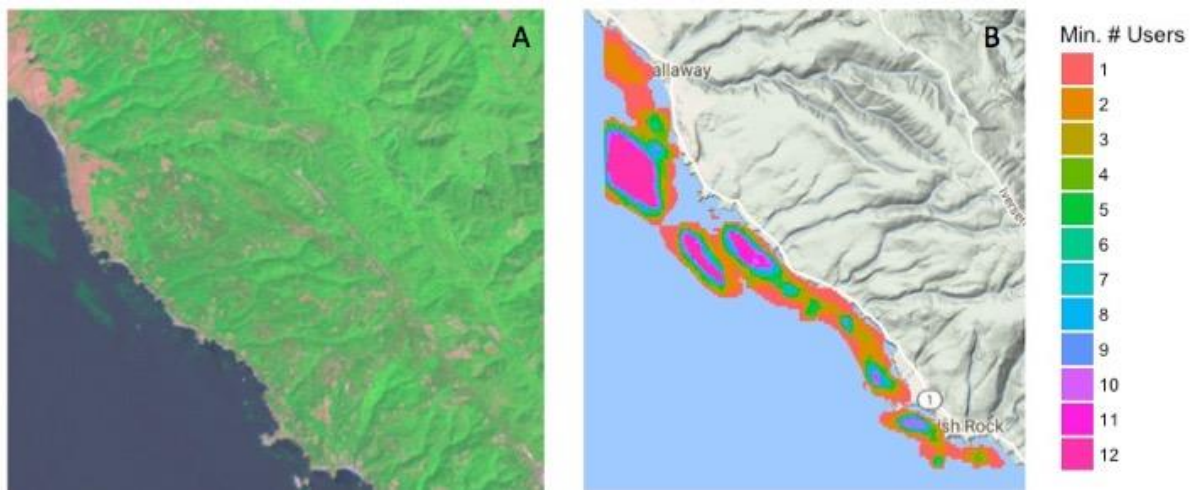

*Figure 5. A) Floating Forest image as presented to users. Note green patches of kelp offshore. B) Heatmap of user consensus thresholds. Landsat-8 image courtesy of the U.S. Geological Survey.*

Our retirement rules were effective at eliminating unwanted images. Figure 4 shows the majority of rejected images to contain little or no kelp. This breakdown also shows a number of rejected images that appear to contain kelp. These are partial or glitched images that received a "bad image" flag and were retired along with the non-kelp images (Figure 5). Partial images are contained in full by neighboring Landsat scenes and thus do not represent missing data.

The consensus approach provides an efficient method to avoid the problem of evaluating citizen scientist expertise. Other attempts to verify and improve citizen science data have suggested weighting a participants contributions based on factors such as length of participation or prior accuracy (Lintott et al., 2008; Whitehill, Ruvolo, Wu, Bergsma, & Movellan, 2009). While this can be effective, it is inefficient as it requires significant overhead on the part of the researcher. Using our system of consensus classifications, the vast majority of data that is collected is utilized.

This is not to say that consensus classification will always be the best choice. Classifier weighting may be required when citizen scientists are responsible for complicated tasks. Consensus classification does require a large number of citizen scientists to evaluate the same thing. In many programs, this might still be impractical. For example, Reef check and other bio-blitz style citizen science projects are typically unable to muster the number of volunteers required to generate a consensus (Reef Check Foundation, 2018). When possible, combining consensus classification with classifier weighting can provide higher accuracy than either method independently (Hutt et al., 2013).

Last, our results alleviate some of the doubts about the effectiveness of web only training that have slowed acceptance of citizen science (Dickinson et al., 2010). Previous work suggests that citizen scientists have higher accuracy when accompanied by professionals (Fitzpatrick, Matthew et al., 2009). Our results demonstrate that citizen scientists can create high quality data despite a small amount of training and only remote communications with experts. While we have no "supervised" data to compare to, the MCC scores of the optimal thresholds provide confirmation that the citizen classifications are accurate relative to expert classifications even in an absence of hands-on expert guidance.

As we move into an unprecedented period of environmental change, it is critical that we consider questions at a global scale. These questions often necessitate datasets derived from long term environmental monitoring efforts, which can be prohibitive for small research teams (Isaac, van Strien, August, de Zeeuw, & Roy, 2014). Citizen science provides a rewarding approach to crowdsource data collection by engaging with volunteers. Despite concerns regarding data quality, publication of results derived from citizen science data has increased substantially over the last 20 years (Follett & Strezov, 2015). We have shown that citizen science data collected via consensus classifications is of adequate quality to use in rigorous scientific analyses. Confidence in data quality is of the utmost importance if citizen science is to be embraced by the scientific community. Consensus classifications are part of an increasingly comprehensive toolkit that can ease quality concerns and increase trust in citizen scientists and their data.

**Acknowledgements**: Thank you to all the citizen scientists whose hard work made this a reality! The authors would like to thank the UMass Boston psychology department stats snack for invaluable advice and NCEAS for the Kelp and Climate Change working group. The authors would also like to recognize the SBC LTER, KEEN, and Temperate Reef Base for making Floating Forests what it is today. Thanks to NASA project 16-CSESP16-0024 for providing funding to support this work.

## References

Bell, T. W., Cavanaugh, K. C., & Siegel, D. A. (2015). Remote monitoring of giant kelp biomass and physiological condition: An evaluation of the potential for the Hyperspectral Infrared Imager (HyspIRI) mission. *Remote Sensing of Environment*, *167*, 218–228. https://doi.org/10.1016/j.rse.2015.05.003

Bird, T. J., Bates, A. E., Lefcheck, J. S., Hill, N. a., Thomson, R. J., Edgar, G. J., … Frusher, S. (2014). Statistical solutions for error and bias in global citizen science datasets. *Biological Conservation*, *173*, 144–154. https://doi.org/10.1016/j.biocon.2013.07.037

Bonney, R., Shirk, J. L., Phillips, T. B., Wiggins, A., Ballard, H. L., Miller-Rushing, A. J., & Parrish, J. K. (2014). Citizen science: Next steps for citizen science. *Science*, *343*(6178), 1436–1437. https://doi.org/10.1126/science.1251554

Bonter, D. N., & Cooper, C. B. (2012). Data validation in citizen science: A case study from Project FeederWatch. *Frontiers in Ecology and the Environment*, *10*(6), 305–307. https://doi.org/10.1890/110273

Boudreau, S. A., & Yan, N. D. (2004). Auditing the Accuracy of a Volunteer-Based Surveillance Program for an Aquatic Invader Bythrephes. *Environmental Monitoring and Assessment*, *91*, 35-3717–26. https://doi.org/10.1023/B

Cavanaugh, K. C., Siegel, D. a., Kinlan, B. P., & Reed, D. C. (2010). Scaling giant kelp field measurements to regional scales using satellite observations. *Marine Ecology Progress Series*, *403*, 13–27. https://doi.org/10.3354/meps08467

Cavanaugh, K. C., Siegel, D. A., Reed, D. C., & Dennison, P. E. (2011). Environmental controls of giant-kelp biomass in the Santa Barbara Channel, California. *Marine Ecology Progress Series*, *429*, 1–17. https://doi.org/10.3354/meps09141

Chicco, D. (2017). Ten quick tips for machine learning in computational biology. *BioData Mining*, *10*(1), 35. https://doi.org/10.1186/s13040-017-0155-3

Cooper, C. B., Dickinson, J., Phillips, T., & Bonney, R. (2007). Citizen science as a tool for conservation in residential ecosystems. *Ecology and Society*, *12*(2). https://doi.org/11

Darwall, W. R. T., & Dulvy, N. K. (1996). An evaluation of the suitability of non-specialist volunteer researchers for coral reef fish surveys. Mafia Island, Tanzania - A case study. *Biological Conservation*, *78*(3), 223–231. https://doi.org/10.1016/0006-3207(95)00147-6

De Solla, S. R., Shirose, L. J., Fernie, K. J., Barrett, G. C., Brousseau, C. S., & Bishop, C. a. (2005). Effect of sampling effort and species detectability on volunteer based anuran monitoring programs. *Biological Conservation*, *121*(4), 585–594. https://doi.org/10.1016/j.biocon.2004.06.018

Delaney, D. G., Sperling, C. D., Adams, C. S., & Leung, B. (2008). Marine invasive species: Validation of citizen science and implications for national monitoring networks. *Biological Invasions*, *10*(1), 117–128. https://doi.org/10.1007/s10530-007-9114-0

Dickinson, J. L., Zuckerberg, B., & Bonter, D. N. (2010). Citizen Science as an Ecological Research Tool: Challenges and Benefits. *Annual Review of Ecology, Evolution, and Systematics*, *41*(1), 149–172. https://doi.org/10.1146/annurev-ecolsys-102209-144636

Fitzpatrick, Matthew, C., Preisser, E. L., Ellison, A. M., Elkinton, J. S., Observer, F., & Link, C. (2009). Observer Bias and the Detection of Low-Density Populations. *Ecological Applications*, *19*(7), 1673–1679.

Foldit. (2018). Foldit. Retrieved from http://fold.it/portal/info/about

Follett, R., & Strezov, V. (2015). An analysis of citizen science based research: Usage and publication patterns. *PLoS ONE*, *10*(11), 1–14. https://doi.org/10.1371/journal.pone.0143687

Greenwood, J. J. D. (2007). Citizens, science and bird conservation. *Journal of Ornithology*, *148*(SUPPL. 1). https://doi.org/10.1007/s10336-007-0239-9

Hutt, H., Everson, R., Grant, M., Love, J., & Littlejohn, G. (2013). How clumpy is my image? *2013 13th UK Workshop on Computational Intelligence (UKCI)*, *19*(6), 136–143. https://doi.org/10.1109/UKCI.2013.6651298

Irwin, A. (2001). Constructing the scientific citizen: science and democracy in the biosciences. *Public Understanding of Science*, *10*, 1–18.

Isaac, N. J. B., van Strien, A. J., August, T. a., de Zeeuw, M. P., & Roy, D. B. (2014). Statistics for citizen science: Extracting signals of change from noisy ecological data. *Methods in Ecology and Evolution*, *5*(10), 1052–1060. https://doi.org/10.1111/2041-210X.12254

Lepczyk, C., Boyle, O., Vargo, T., Gould, P., Jordan, R., & Al., E. (2009). Citizen science in ecology: the intersection of research and education. In *Bulletin of the Ecological Society of America* (pp. 308–317).

Lewandowski, E., & Specht, H. (2015). Influence of volunteer and project characteristics on data quality of biological surveys. *Conservation Biology*, *29*(3), 713–723. https://doi.org/10.1111/cobi.12481

Lintott, C. J., Schawinski, K., Slosar, A., Land, K., Bamford, S., Thomas, D., … Vandenberg, J. (2008). Galaxy Zoo: Morphologies derived from visual inspection of galaxies from the Sloan Digital Sky Survey. *Monthly Notices of the Royal Astronomical Society*, *389*(3), 1179–1189. https://doi.org/10.1111/j.1365-2966.2008.13689.x

National Audubon Society. (2018). Christmas Bird Count. Retrieved from http://www.audubon.org/conservation/science/christmas-bird-count

National Audubon Society, & The Cornell Lab of Ornithology. (2018). eBird. Retrieved from http://ebird.org/content/ebird/

North American Butterfly Association. (2018). North American Butterfly Association-Butterfly Counts. Retrieved March 1, 2018, from http://www.naba.org/publications.html

Reef Check Foundation. (2018). Reef Check Foundation: What We Do. Retrieved January 1, 2018, from http://www.reefcheck.org/what-we-do

Ricciardi, B. Y. A., Steiner, W. W. M., Mack, R. N., & Simberloff, D. (2000). Toward a Global Information System for Invasive Species. *Bioscience*, *50*(3), 239–244.

SciStarter. (2018). Project Finder. Retrieved January 1, 2018, from https://scistarter.com/finder?phrase=&lat=&lng=&location_text=&activity=&topic=#view-projects

Silvertown, J. (2009). A new dawn for citizen science. *Trends Ecol. Evol.*, *24*(9), 467–471. https://doi.org/10.1016/j.tree.2009.03.017

Swanson, A., Kosmala, M., Lintott, C., Simpson, R., Smith, A., & Packer, C. (2015). Snapshot Serengeti, high-frequency annotated camera trap images of 40 mammalian species in an African savanna. *Scientific Data*, *2*, 1–14. https://doi.org/10.1038/sdata.2015.26

The Cornell Lab of Ornithology. (2018a). FeederWatch. Retrieved from http://feederwatch.org/

The Cornell Lab of Ornithology. (2018b). NestWatch. Retrieved from https://nestwatch.org/

USGS Patuxent Wildlife Research Center. (2018). North American Breeding Bird Survey. Retrieved from http://www.pwrc.usgs.gov/bbs/

Vihinen, M. (2012). How to evaluate performance of prediction methods? Measures and their interpretation in variation effect analysis. *BMC Genomics*, *13*(Suppl 4), S2. https://doi.org/10.1186/1471-2164-13-S4-S2

Whitehill, J., Ruvolo, P., Wu, T., Bergsma, J., & Movellan, J. (2009). Whose Vote Should Count More: Optimal Integration of Labels from Labelers of Unknown Expertise. *Advances in Neural Information Processing Systems*, *22*(1), 1–9.

Willett, K. W., Lintott, C. J., Bamford, S. P., Masters, K. L., Simmons, B. D., Casteels, K. R. V, … Thomas, D. (2013). Galaxy zoo 2: Detailed morphological classifications for 304 122 galaxies from the sloan digital sky survey. *Monthly Notices of the Royal Astronomical Society*, *435*(4), 2835–2860. https://doi.org/10.1093/mnras/stt1458

Zooniverse. (2018). Zooniverse. Retrieved January 2, 2018, from https://www.zooniverse.org/about

**Supplemental material:**

| Coefficient | Estimate | Std. Error | t value | *p*-value |
|---|---|---|---|---|
| Intercept | 0.262 | 0.038 | 6.860 | <0.001 |
| User threshold linear term | 0.058 | 0.019 | 3.061 | 0.002 |
| User threshold squared term | -0.006 | 0.002 | -3.752 | 0.000 |
| Landsat 7 | 0.113 | 0.063 | 1.793 | 0.073 |
| Landsat 8 | 0.262 | 0.044 | 5.952 | 0.000 |
| Spring | -0.074 | 0.048 | -1.542 | 0.124 |
| Summer | -0.050 | 0.052 | -0.958 | 0.338 |
| Winter | 0.033 | 0.051 | 0.639 | 0.523 |
| User threshold : Landsat 7 | -0.031 | 0.028 | -1.114 | 0.266 |
| User threshold $^2$ : Landsat 7 | 0.001 | 0.002 | 0.404 | 0.686 |
| User threshold : Landsat 8 | -0.013 | 0.022 | -0.585 | 0.559 |
| User threshold $^2$ : Landsat 8 | 0.002 | 0.002 | 0.807 | 0.420 |
| User threshold : Spring | 0.025 | 0.023 | 1.060 | 0.289 |
| User threshold $^2$: Spring | -0.002 | 0.002 | -0.757 | 0.449 |
| User threshold : Summer | 0.013 | 0.027 | 0.506 | 0.613 |
| User threshold $^2$ : Summer | 0.000 | 0.003 | -0.110 | 0.912 |
| User threshold : Winter | -0.007 | 0.024 | -0.282 | 0.778 |
| User threshold $^2$ : Winter | 0.000 | 0.002 | 0.210 | 0.834 |

Supplemental table 1: coefficients from linear model